# Realistic Approach towards Quantitative Analysis and Simulation of EEHC-Based Routing for Wireless Sensor Networks

Manju Sharma and Lalit Awasthi

National Institute of Technology
Hamirpur, HP-177005, India

**Abstract**
This paper presents the realistic approach towards the quantitative analysis and simulation of Energy Efficient Hierarchical Cluster (EEHC)-based routing for wireless sensor networks. Here the efforts have been done to combine analytical hardware model with the modified EEHC-based routing model. The dependence of various performance metrics like: optimum number of clusters, Energy Consumption, and Energy consumed per round etc. based on analytical hardware sensor model and EEHC model has been presented.

## 1. Introduction

The basic element of a wireless sensor networks (WSN) is the sensor node, where it is consisted of three main functional components that separately deliver sensory, communication and processing capabilities. Because of the difficulty and cost of sensor node replacement in face of battery drainage or system failure, hardware reliability is also another major concern in WSN design. Therefore keeping in realistic facts and operational conditions the focus of this paper is report the realistic quantitative analysis EEHC-based routing for WSN.

Many energy-efficient routing protocols are designed based on the clustering structure. The randomized clustering algorithm to organize sensors into clusters in a WSN was proposed Bandyopadhyay and Coyle [1]. Further the computation of the optimal probability of becoming a cluster head was presented. In [2], Moscibroda and Wattenhofer defined the maximum cluster-lifetime problem, and they proposed distributed, randomized algorithms that approximate the optimal solution to maximize the lifetime of dominating sets on WSNs.

Low-energy adaptive clustering hierarchy (LEACH) [3] is one of the first hierarchal routing approaches for WSNs. It is a well-known clustering protocol for wireless sensor networks and most of the clustering algorithms are based on this algorithm. This protocol uses only two layers for communication. LEACH includes distributed cluster formation, local processing to reduce global communication, and randomized rotation of cluster-heads. In literature it has been reported that LEACH performs over a factor of 7 reductions in energy dissipation compared to flat base routing algorithm such as direct diffusion [4]. But the main problem with LEACH protocol lies in the random selection of cluster heads. There exists a probability that the cluster heads formed are unbalanced and may remain in one part of the network making some part of the network unreachable.

As an extension of LEACH [3, 5-7], the proposed protocol introduces a head set for the control and management of clusters. Although S-MAC [8] divides the network into virtual clusters, the proposed protocol divides the network into a few real clusters that are managed by a virtual cluster-head. The results reported in this paper are the efforts towards the realistic approach by extending of work reported in [7, 9]. The derivations for transceiver power consumption modeled in [9] and [7] are considered to report modified quantitative analysis for the EEHC based routing for WSN.

The rest of this paper is organized as follows. After mentioning the introduction in section 1, the related work is discussed in section 2. In Section 3, we have described the quantitative analysis for EEHC-based WSN. In section 4 the performances have been evaluated of EEHC via simulations. Finally, Section 5 concludes the paper and future work is pointed out.

## 2. Related Work

The cluster-based structure is widely acknowledged as an energy conserving way to facilitate the network self-organization as well as approximate the global optimization of a WSN [10], in particular for periodic reporting applications, where data from many sources converge into one data sink. Clustering based routing performs very efficiently when applied to homogeneous WSNs.

Mhatre and Rosenberg [11] studied the case of multi-hop routing within each cluster, which is called M-LEACH. In M-LEACH, only powerful nodes can become the cluster-heads. Energy Efficient Cluster Scheme (EECS) [12] elects the cluster-heads with more residual energy through local radio communication. But on the other hand, it increases the requirement of global knowledge about the distances between the cluster-heads and the base station. A new adaptive strategy is proposed known as LEACH-B in [13] to choose cluster-heads and to vary their election frequency according to the dissipated energy. However, the simulation results divulge that there is some degree of improvement using LEACH-B. Moreover, an improved scheme of LEACH was proposed, named LEACH-C [5]. In LEACH-C, a centralized algorithm at the base station makes cluster formation. However, LEACH-C is not feasible for larger networks because nodes far away from the base station will have problem sending their states to the base station and as the role of cluster heads rotates so every time the far nodes will not reach the base station in quick time increasing the latency and delay.

Further, the clustering protocol known as LEACH-E was proposed by Heinzelman et.al. in [14]. In this protocol it is proposed to elect the cluster-heads according to the energy left in each node. The drawback of LEACH-E is that it requires the assistance of routing protocol, which should allow each node to know the total energy of network. Stable Election Protocol (SEP) [15] was developed for the two-level heterogeneous networks. SEP performs poorly in multi-level heterogeneous networks and when heterogeneity is a result of operation of the sensor network. Distributed Energy-Efficient Clustering (DEEC) [16], which is dedicatedly designed for energy heterogeneous scenarios, where nodes are initialized at various energy levels. However neither of them assures the selection of energy-rich cluster heads, or the evenness of cluster head dispersion. Decentralized Energy Efficient clustering Propagation (DEEP) [17] prevents cluster heads from being too close to each other, but ignores cluster head's energy qualifications.

In [18], Lindsey et al. proposed Power-Efficient Gathering in Sensor Information Systems (PEGASIS). PEGASIS makes a communication chain using a Traveling Sales Person heuristic. Each node only communicates with two close neighbors along the communication chain. Only a single designated node gathers data from other nodes and transmits the aggregated data to the sink node. Tan and Korpeoglu in [19], proposed two new algorithms under the name Power Efficient Data Gathering and Aggregation Protocol (PEDAP), which are near optimal minimum spanning tree based wireless routing scheme. The performance of the PEDAP was compared with LEACH and PEGASIS, and showed a slightly better network lifetime than PEGASIS.

Manjeshwar et. al. developed a protocol called Adaptive Periodic Threshold-sensitive Energy Efficient sensor Network Protocol (APTEEN) [20] that uses an enhanced TDMA schedule to efficiently incorporate query handling. APTEEN provides a combination of proactive and reactive policies. In the Hierarchy-Based Anycast Routing (HAR) Protocol for WSNs [21], the sink constructs a hierarchical tree by sending packets to discover each node's own child nodes in turn. The drawback of HAR is that it sends and receives too many packets in the network, expending much energy. Therefore a novel Hierarchy-Based Multipath Routing Protocol (HMRP) is proposed by Wang et. al. [22] to overcome the defects of HAR.

In [23], Yang et al. proposed a new routing scheme; called Shortest Hop Routing Tree (SHORT), to achieve higher energy efficiency, network lifetime, and more throughput than PEGASIS, and PEDAP-PA protocols. This scheme used the centralized algorithms and required the powerful base station. Hybrid Energy-Efficient Distributed algorithm (HEED) [24] is capable of setting up a cluster head in the center of a dense area irrespective of network topologies. The probability for each node to become a tentative cluster head depends on its residual energy, and all the tentative heads in which are competing for becoming the final cluster heads.

Time-driven clustering schemes have been recently developed. Clustering Algorithm via Waiting Timer (CAWT) [25] still relies on the local messages exchange to calculate the timer. Backoff strategy clustering scheme [26], which is advance in selecting energetic nodes as cluster heads and eliminating huddling cluster heads, suggests a novel way to configure the timer. In [27], Li et. al. proposed Energy–Efficient Unequal Clustering (EEUC) protocol. It divides the nodes into clusters of unequal size. An overhead free fully distributed clustering scheme, called Slotted Waiting period Energy-Efficient Time Driven clustering algorithm (SWEET) is proposed in [28]. Though SWEET is an overhead-free method, it performs even better than some representative clustering schemes in extending system lifetime and enlarging network data capacity.

## 3. Proposed Model

The proposed routing scheme is based on the fact that the energy consumed to send a message to a distant node is far greater than the energy needed for a short range transmission. Here a simple radio frequency (RF) transceiver model (Fig. 1) has been considered that is connected to the sensing and processing unit of the sensor node. Data to be transmitted will first pass through the digital-to-analog converter (DAC) and low-pass filter to prepare for up-conversion at the mixer with carrier signal generated by the frequency synthesizer. The subsequent modulated signal will be transmitted by the power amplifier (PA) over the wireless channel.

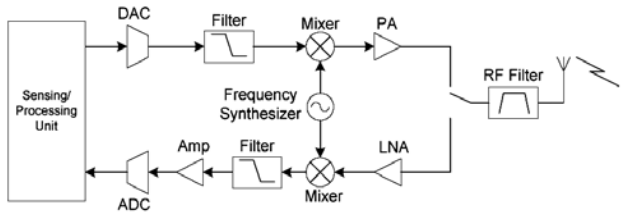

Fig. 1 A simple RF transceiver model of sensor node

On the receive path, incoming signals will first be amplified by the low noise amplifier (LNA), and then demodulated and processed through a series of intermediate frequency (IF) and base-band filters and amplifiers (not explicitly shown in Fig. 1 [9]. In the end, digital data are recovered by the analog-to-digital converter (ADC) and then forwarded to the processing unit for further decoding.

We have extended the LEACH protocol by using a head-set instead of a cluster head [7] in combination with transceiver model of sensor node [9]. In other words, during each election, a head-set that consists of several nodes is selected. The members of a head-set are responsible for transmitting messages to the distant base station. At one time, only one member of the head-set is active and the remaining head-set members are in sleep mode. The above communication stages are illustrated in Fig. 2.

## 4. Quantitative Analysis

The following derivation for transceiver power consumption is modeled after those given in [9], and has been used later for the analysis. On the transmission path, the total power consumption, $P_{Tx}$, can be written as:

$$P_{Tx} = P_{PA} + P_E \qquad (1)$$

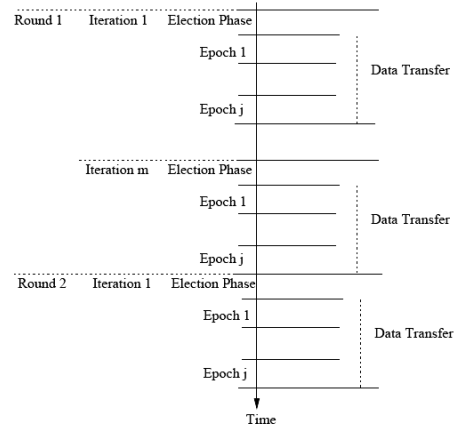

Fig. 2 Communication stages in a cluster of a wireless sensor network [7]

Where $P_{PA}$ is the amount of power consumed by the $P_A$ alone, and $P_E$ is the amount of power collectively consumed by the other electronic components such as the mixer, frequency synthesizer, DAC, and various filters. Determining the exact values for both $P_{PA}$ and $P_E$ would depend on RF component design and device technology, which is beyond the scope of this research; though a simple approximation would suffice in the current work. While $P_E$ is generally treated as a constant under various operating conditions, $P_{PA}$ can be further broken down into the following terms:

$$P_{PA} = \frac{P_{RxSi}\left(\frac{1}{L_o}\right)\left(\frac{R_{com}}{d_o}\right)^\alpha}{G_{Tx} G_{Rx} \eta} \qquad (2)$$

Where $P_{RxSi}$ denotes the receiver sensitivity in Watts, $L_o$ is the path loss attenuation at $d_o$ metres, $R_{com}$ refers to the distance between the transmitter and the receiver in metres, $\alpha$ is the path loss exponent, $G_{Tx}$ and $G_{Rx}$ represent transmit and receive antenna gains, respectively, and $\eta$ stands for PA efficiency. In turn, the receiver sensitivity $P_{RxSi}$ can be rewritten as:

$$P_{RxSi} = \left(\frac{S}{N}\right)_{Rx} \cdot NF_{Rx} \cdot N_o \cdot BW \qquad (3)$$

Where $\left(\frac{S}{N}\right)_{Rx}$ is the minimum signal-to-noise ratio that provides an acceptable $\left(\frac{E_b}{N_o}\right)$ level at the receiver, $N_{FRx}$ is the noise factor at the receiver, $No$ is the thermal noise floor in a 1 Hz bandwidth (in W/Hz or J), and $BW$ is the channel noise bandwidth (in Hz).

Let:
$$\frac{1}{L_o} = \left(\frac{4\pi}{w}\right)^\alpha, \quad d_o = 0.1 \qquad (4)$$

Where $w$ is the wavelength of the carrier frequency in meters. Substituting Eq.s (3) and (4) into (2), $P_{PA}$ becomes:

$$P_{PA} = \frac{\left(\frac{S}{N}\right)_{Rx} \cdot NF_{Rx} \cdot N_o \cdot BW \cdot \left(\frac{4\pi}{w}\right)^\alpha \cdot 10^\alpha \cdot d^\alpha}{G_{Tx} G_{Rx} \eta} \quad (5)$$

In terms of $\left(\frac{E_b}{N_o}\right)$, $P_{PA}$ can be represented as:

$$P_{PA} = \frac{\left(\frac{E_b}{N_o}\right)_{Rx} \cdot R_{Tx} \cdot NF_{Rx} \cdot N_o \cdot BW \cdot \left(\frac{4\pi}{w}\right)^\alpha \cdot 10^\alpha \cdot d^\alpha}{G_{Tx} G_{Rx} \eta} \quad (6)$$

On the reception path, the total power consumption, $P_{Rx}$, depends on the power consumption of the LNA, mixer, frequency synthesizer, IF amplifiers, filters, and ADC.

Generally the sensor model doesn't specify the transmitting or receiving one bit. Nonetheless, the platform uses transmission rate of 1 Mbps ($R_{bits}$) or time to send one bit is 1 μs, so one can calculate the energy required for transmitting one bit, following a method based in the approach presented by Hill et.al in [29]. The energy used in transmitting or receiving one bit and is found by using the power value and can be further derived as:

$$\varepsilon_s = P_{PA} \times time = \frac{P_{PA}}{R_{bit}} \quad (7)$$

Where $R_{bit}$ is the raw bit rate, time in seconds and power in Watts.

$$\varepsilon_s = \frac{\left(\frac{S}{N}\right)_{Rx} \cdot NF_{Rx} \cdot N_o \cdot BW \cdot \left(\frac{4\pi}{w}\right)^\alpha \cdot 10^\alpha \cdot d^\alpha}{G_{Tx} G_{Rx} \eta R_{bit}} \quad \text{or in}$$

terms of $\left(\frac{E_b}{N_o}\right)$ can be represented as:

$$\varepsilon_s = \frac{\left(\frac{E_b}{N_o}\right)_{Rx} \cdot R_{Tx} \cdot NF_{Rx} \cdot N_o \cdot BW \cdot \left(\frac{4\pi}{w}\right)^\alpha \cdot 10^\alpha \cdot d^\alpha}{G_{Tx} G_{Rx} \eta R_{bit}} \quad (8)$$

The radio communication and energy consumption described in [7] is adopted: for short distance transmission, such as intra-cluster communication, the energy consumed by a transmitting amplifier is proportional to $d^2$ and for long distance transmission, such as inter-cluster communication, the energy consumption is proportional to $d^4$. Using the given radio and energy consumption models, the energy consumed in transmitting one message among cluster heads for a distance $d$ is given by

$$E_T = lE_e + l\varepsilon_l d^4 \quad (9)$$

Similarly, the energy consumed when the senor node works as a regular (member) node, that is, the energy consumed in transmitting a massage within a cluster for a short distance d, is given by

$$E_T = lE_e + l\varepsilon_s d^2 \quad (10)$$

Moreover, the energy consumed to receive the l-bit message is given by:

$$E_T = lE_e + lE_{BF} \quad (11)$$

Eq. 11 includes the cost of beam forming approach that reduces energy consumption. The constants used in the radio model are given in Table 1.

**Table 1:** Parameters values used for radio communication model and other for quantitative analysis

| Description | Symbol | Value |
|---|---|---|
| Energy consumed by the amplifier to transmit at a longer distance | $\varepsilon_l$ | 0.0013 pJ/bit/m$^4$ |
| Energy consumed in the electronics circuit to transmit or receive the signal | $E_e$ | 50 nJ/bit |
| Energy consumed for beam forming | $E_{BF}$ | 5 nJ/bit |
| Noise factor at the receiver | $NF_{Rx}$ | 11 dB (12.589) |
| Minimum signal-to-noise ratio that provides an acceptable $E_b/N_o$ level at the receiver | $\left(\frac{S}{N}\right)_{Rx}$ | 10 dB (10) |
| Thermal noise floor in a 1 Hz bandwidth (in W/Hz or J) | $N_o$ | -173.8 dBm/Hz (4.17 X 10$^{-21}$J) |
| Wavelength of the carrier frequency in meters | $w$ | 0.328 m (for 915 MHz) 0.125 m (for 2.4 GHz) |
| Transmit and receive antenna gains | $G_{Tx} G_{Rx}$ | -20 dBi (0.01) |
| Stands for PA efficiency | $\eta$ | 0.2 |
| Path loss exponent | $\alpha$ | 2 |
| Channel noise bandwidth (in Hz) | $BW$ | 1 bit/Hz X $B_{Tx}$ |
| Amount of power collectively consumed by the other electronic components such as the mixer, frequency synthesizer, DAC, and various filters | $P_E$ | 3.63 mW |

There is uniform distribution of clusters and each cluster contains $n/k$ nodes. For a sensor network of $n$

nodes, the optimal number of clusters is given as *k*. It is assumed that:

- The quantitative analysis is based on the radio model for shorter distance Eq. 9.
- All nodes are at the same energy level at the beginning.
- The amount of consumed energy is same for all the clusters.
- At the start of the election phase, the base station randomly selects a given number of cluster heads.
- First, the cluster heads broadcast messages to all the sensors in their neighborhood.
- Second, the sensors receive messages from one or more cluster heads and choose their cluster head using the received signal strength.
- Third, the sensors transmit their decision to their corresponding cluster heads.
- Fourth, the cluster heads receive messages from their sensor nodes and remember their corresponding nodes.
- For each cluster, the corresponding cluster head chooses a set of *m* associates, based on signal analysis.

4.1 Election phase

Using Eq. 8, 10 and Eq. 11, the energy consumed by a cluster head is estimated as follows:

$$E_{CH-elec} = \left\{ l E_e + l \frac{\left(\frac{E_b}{N_o}\right)_{Rx} \cdot R_{Tx} \cdot NF_{Rx} \cdot N_o \cdot BW \cdot \left(\frac{4\pi}{w}\right)^\alpha \cdot 10^\alpha \cdot d^\alpha}{G_{Tx} G_{Rx} \eta R_{bit}} \right\}$$
$$+ \left\{ \left(\frac{n}{k} - 1\right) l \left(E_e + E_{BF}\right) \right\} \quad (12)$$

Similarly the energy consumed by non-cluster head sensor nodes is estimated as follows:

$$E_{non-CH-elec} = \left\{ l E_e + l \frac{\left(\frac{E_b}{N_o}\right)_{Rx} \cdot R_{Tx} \cdot NF_{Rx} \cdot N_o \cdot BW \cdot \left(\frac{4\pi}{w}\right)^\alpha \cdot 10^\alpha \cdot d^\alpha}{G_{Tx} G_{Rx} \eta R_{bit}} \right\}$$
$$+ \left\{ k l \left(E_e + E_{BF}\right) \right\} \quad (13)$$

4.2 Data transfer phase

As mentioned in [7] during data transfer phase, the nodes transmit messages to their cluster head and cluster heads transmit an aggregated messages to a distant base station. The energy consumed by a cluster head is as follows:

$$E_{CH/frame} = \left\{ l E_e + l \varepsilon_l d^4 \right\} + \left\{ \left(\frac{n}{k} - m\right) l \left(E_e + E_{BF}\right) \right\} \quad (14)$$

The first part of Eq. 14 shows the energy consumed to transmit a message to the distant base station. The second part of Eq. 14 shows the energy consumed to receive messages from the remaining $\left(\frac{n}{k} - m\right)$ nodes that are not part of the head-set.

The energy, $E_{non-CH/frame}$, consumed by a non-cluster head node to transmit the sensor data to the cluster head is given below:

$$E_{non-CH/frame} = l E_e + l \frac{\left(\frac{E_b}{N_o}\right)_{Rx} \cdot R_{Tx} \cdot NF_{Rx} \cdot N_o \cdot BW \cdot \left(\frac{4\pi}{w}\right)^\alpha \cdot 10^\alpha \cdot d^\alpha}{G_{Tx} G_{Rx} \eta R_{bit}} \quad (15)$$

The energy consumptions in a data transfer stage of each cluster are as follows:

$$E_{CH-data} = f_1 N_f E_{CH/frame} \quad (16)$$
$$E_{non-CH-data} = f_2 N_f E_{non-CH/frame} \quad (17)$$

Where $f_1 = \left(\frac{1}{\frac{n}{k} - m + 1}\right) \cdot \frac{1}{k}$ & $f_2 = \left(\frac{\frac{n}{k} - m}{\frac{n}{k} - m + 1}\right) \cdot \frac{1}{k}$

4.3 Starting energy for one round

The start energy, $E_{start}$, is energy of a sensor node at the initial start time. This energy should be sufficient for at least one round. In one round, a node becomes a member of head-set for one time and a non-cluster head for $\left(\frac{n}{km} - 1\right)$ times. An estimation of $E_{start}$ is given below:

$$E_{start} = \frac{E_{CH-elec} + E_{non-CH-elec}}{m} + \frac{N_f}{m} \left( f_1 E_{CH/frame} + f_2 E_{non-CH/frame} \right) \quad (18)$$

and from this Eq. $N_f$ can be derived as:

$$N_f = \frac{m E_{start} - E_{CH-elec} - E_{non-CH-elec}}{f_1 E_{CH/frame} + f_2 E_{non-CH/frame}} \quad (19)$$

Here it has been assumed that there are *k* clusters and *n* nodes. In each iteration, *m* nodes are elected for each cluster. Thus, in each iteration *k m* nodes are elected as members of head-sets. The number of iterations required for all *n* nodes to be elected is $\left(\frac{n}{km}\right)$, which is the number of iterations required in one round. Moreover, an iteration consists of an election phase and a data transfer stage.

4.4 Optimum number of clusters

Following [7] and [9] the optimum number of *k* for minimum consumed energy can be determined as follows:

$$k = \sqrt{\frac{n}{2\pi}} \cdot \sqrt{\frac{\left(\frac{E_b}{N_o}\right)_{Rx} \cdot R_{Tx} \cdot NF_{Rx} \cdot N_o \cdot BW \cdot \left(\frac{4\pi}{w}\right)^\alpha \cdot 10^\alpha \cdot d^\alpha}{G_{Tx} G_{Rx} \eta R_{bit}}{\varepsilon_1 d^4 - (2m-1)E_e - m E_e - m E_{BF}}} \cdot M \quad (20)$$

## 4. Results and discussion

Fig. 3 illustrate the graph that indicates the variation of optimum number of clusters with respect to the head set size where it is assumed that the base station is at the distance of 150 m at different number of nodes. As the graph indicate that the optimum value for head set size observed to be six, while number of clusters requirement in case of n = 1000, 1500, 2000 is 14, 17 and 20 respectively. Further for hierarchical cluster based routing of wireless sensor network for a given head set size it is easy to determine the maximum number of clusters from this graph.

Fig. 3(b) shows the graph that illustrates the variation in number of optimum number of clusters at different values of head set size in case of each iteration at different values of nodes to be elected for each cluster. The graph predict that for each optimum head set size of 6 the requirements of number of clusters increases to 14, 29 and 44 respectively in case of *m* = 100, 200 and 300. The results conclude that the number of clusters requirement increases as there is increase in number of m.

Fig. 4 depicts that the average time to compute each iteration versus number of clusters and head set size at different values of nodes. The figure clearly indicates that as the number of head set size increases and number of clusters (*k*) decreases and there is significant increase in time for each iteration in both cases i.e. *n* = 1000 & 2000. Further in both cases results indicate that there is significant decrease in time for single iteration if the numbers of clusters have been increased.

Fig. 5 indicates the energy consumption in data transfer stage each cluster head node and non-cluster head node at different diameter and head set size. The result clearly shows that there is increase in energy consumption for non-cluster head node if the cluster head size is 10 to 15 and otherwise there is exponential decrease. On the other hand the energy consumption of cluster head node is highest when the head set size is maximum i.e. 20 and there is exponential decrease in energy consumption for cluster head node as head set size reduces whereas trend is opposite in case of non-cluster head.

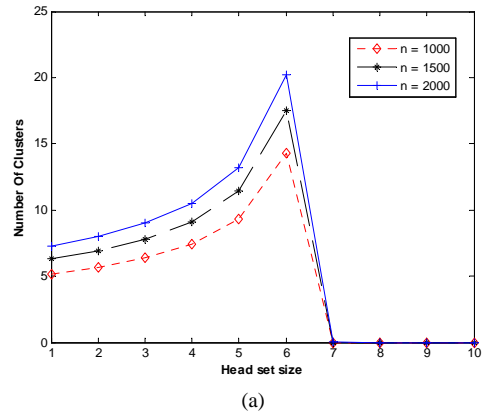

(a)

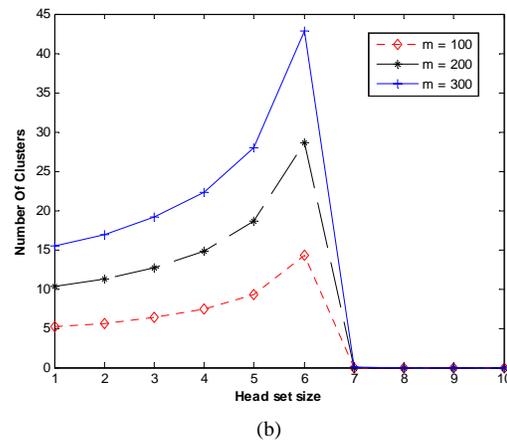

(b)

Fig. 3 (a) Maximum optimum number of clusters at different number of nodes *n* =1000, 1500 and 2000 and (b) Maximum optimum number of clusters for corresponding cluster head chooses a set of m associates (*m* =100, 200 and 300).

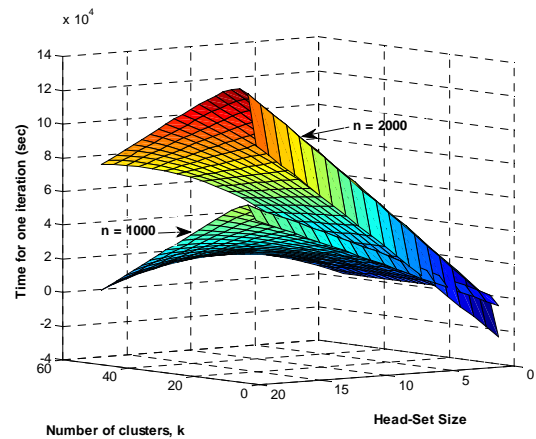

Fig. 4 Number of cluster and headset size for one iteration.

Fig. 6 illustrate the number of clusters and head set size at different SNR values of sensor nodes. As indicated in figure 1 earlier that the head set size is optimum at 6.

This figure illustrate that at optimum value of head set size the number of cluster requirement vary and is 11,15 and 19 at SNR values of 10, 20 and 30 dB respectively. These results reveal that at the optimum head set size the requirement of number of clusters is dependent on SNR value of sensor node.

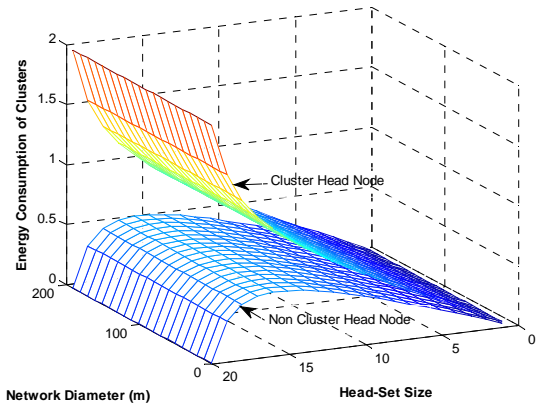

Fig. 5 Energy Consumption in data transfer stage of each cluster head node and non-cluster head node

Fig. 7 shows the illustration of maximum optimum number of clusters at different efficiency values of sensor node power amplifier for the different head set size. Similar to figure 1 & 4 the figure 7 clearly depicts that the requirement of number of clusters at optimum value of head set size of 6 vary as there is change in efficiency of PA. The number of cluster requirement is indirectly dependent upon the efficiency of sensor node power amplifier. At optimum head set size higher the efficiency, less is the requirement for number of clusters. Results ascertain that at efficiency the values of 0.6, 0.4 and 0.2 the number of cluster requirement is 9, 11 and 15.8 respectively.

The energy consumption at the maximum optimum number of clusters at head-set size of m = 1 with distance d = 150 and 400 m from the base station has been indicated in figure 8. As obvious and it clearly shows the dependence of energy requirement on distance between the sensor nodes.

Energy consumption at maximum optimum number of clusters at head set size of m = 1 with different data frames transmitted in one iteration has been indicated in Fig. 9 (a & b). This has been shown in Fig. 9 that the energy consumption is dependent upon the number of data frames transmitted in one iteration. Higher is the data frames, higher is the consumption. While Fig. 9 (b) indicate that there is significant decrease in energy consumption if the head set size enhanced to 3.

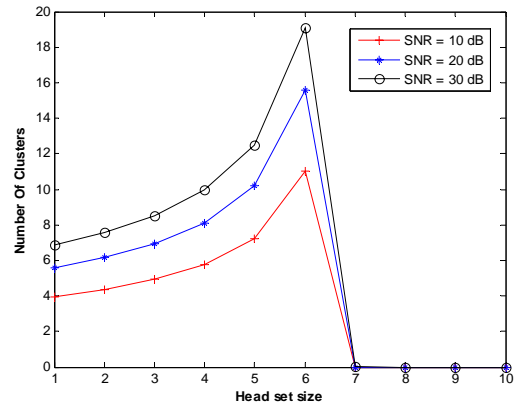

Fig. 6 Maximum optimum number of clusters at different SNR values of sensor node.

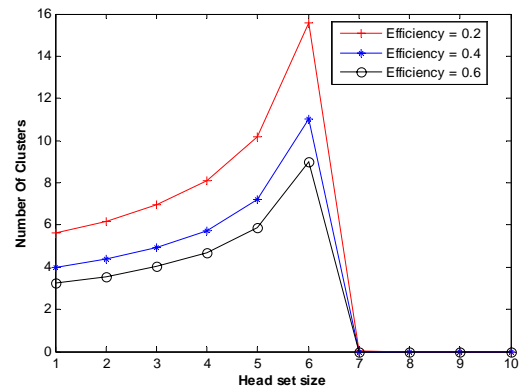

Fig. 7 Maximum optimum number of clusters at different efficiency of values of sensor node PA.

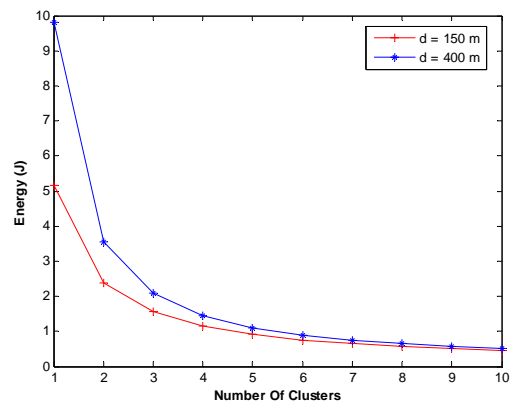

Fig. 8 Energy consumption at the maximum optimum number of clusters at Head-set size of m = 1 with distance d = 150 and 400 m from the base station

Fig. 10 illustrate the energy consumption at the maximum optimum number of clusters for different head set size in case of clusters and non-clusters head node for single iteration. The figure illustrate that the energy consumption is higher in case of clusters head node if head set size is 1, while trends are reverse if the head set size is 3 (Fig. 10(b)).

Fig. 11 shows the energy consumed per round/iteration with respect number of clusters at different distance from the base station. Results establish that the starting energy requirement is dependent on number of clusters and the network diameter from the base station. Less starting energy is required if number of clusters are high and d is less otherwise it is opposite.

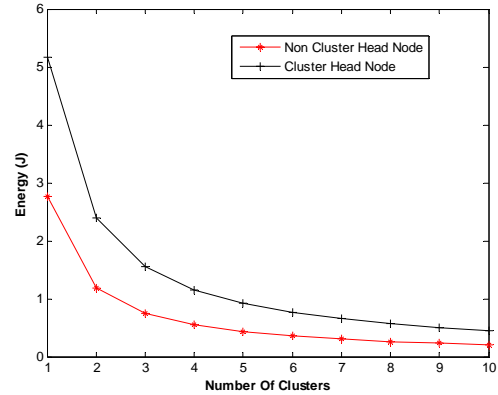

(a)

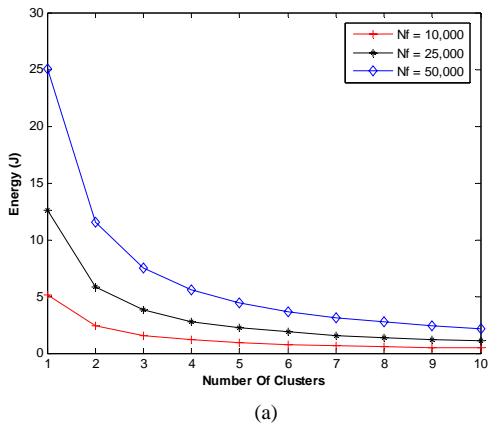

(a)

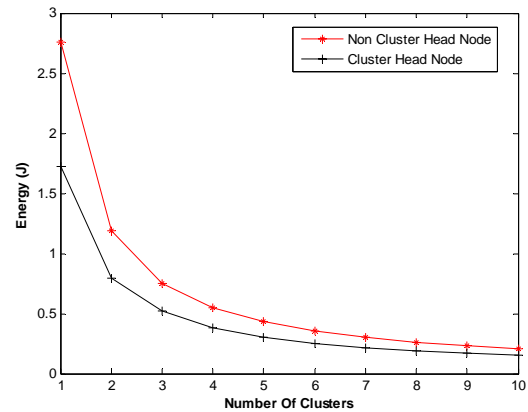

(b)

Fig. 10 Energy consumption at the maximum optimum number of clusters at Head-set size of (a) m = 1 and (b) m = 3 with data frames Nf = 10000 transmitted in one iteration for Non Cluster Head and Cluster Head node

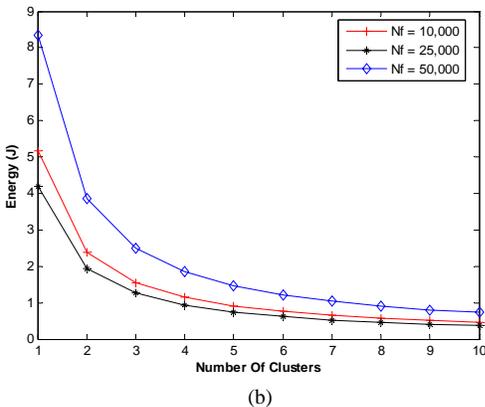

(b)

Fig. 9 Energy consumption at the maximum optimum number of clusters at Head-set size of (a) m = 1 and (b) m = 3 with data frames Nf = 10000, 25000 and 50000 transmitted in one iteration

Fig. 12 indicates the energy consumed per round with respect to number of clusters and network diameter at different values of data frames. It has been observed that the starting/initial energy requirement is dependent upon the data frames used. It must be kept highest when number of clusters are less and network diameter is more.

The energy consumed per round/ iteration with respect to number of clusters and network diameter at cluster head set size of 3 are indicated in fig. 13.

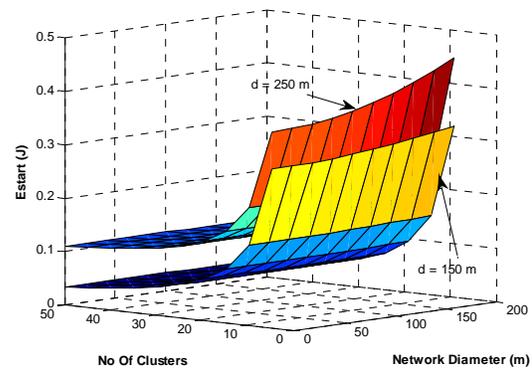

Fig. 11 Energy consumed per round with respect to number of clusters at different values of d from the base station.

Fig. 13 clearly indicates the energy consumed per round at start for clusters and non clusters head. It has been observed that the energy consumption is non clusters head is exponential, while it is linear in case of cluster head with respect to number of clusters and network diameter.

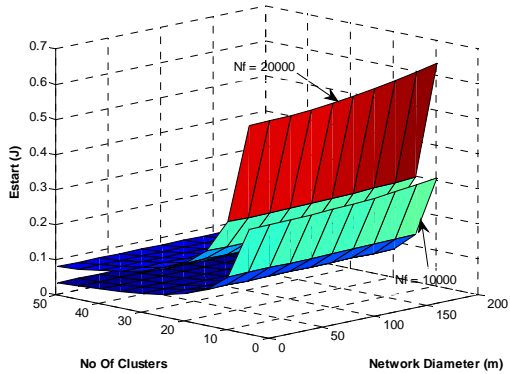

Fig. 12 Energy consumed per round with respect to number of clusters and diameter at different values of Nf

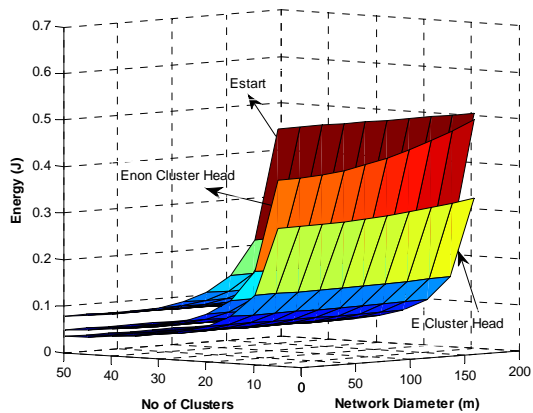

Fig. 13 Energy consumed per round with respect to number of clusters and network diameter at cluster head size m = 3.

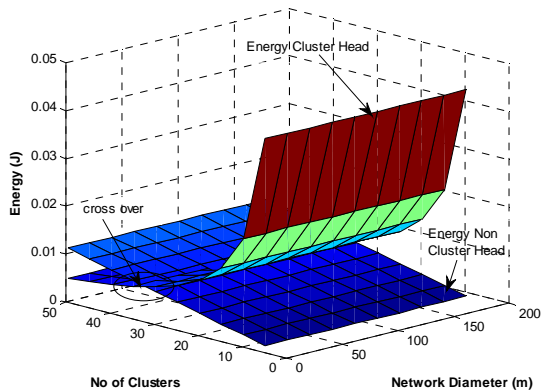

Fig. 14 Energy consumed per round with respect to number of clusters and Non Cluster Head under similar conditions at cluster head size m = 3

Fig. 14 illustrate the energy consumption per round/ iteration with respect to the number of clusters and non-clusters head at head set size of 3. Here highest energy consumption shown in case of cluster head size of 6, while it is minimum in case of non- cluster head. It is interesting to note that there is cross over point for energy consumption between cluster head and non cluster head at 45 numbers of clusters. Here the energy requirement of non-cluster head rises over cluster head.

## 5. Conclusions

This paper presents the observations obtained from the combination of realistic analytical wireless sensor model and EEHC-based routing for wireless sensor networks. The results predict that for each optimum head set size of 6 the requirements of number of clusters increases to 14, 29 and 44 respectively in case of $m$ = 100, 200 and 300. Further the results illustrate that at optimum value of head set size the number of cluster requirement vary and is 11,15 and 19 at SNR value of 10, 20 and 30 dB respectively. These results divulge that at the optimum head set size the requirement of number of clusters is dependent on SNR value of sensor node. At optimum head set size higher the efficiency, less is the requirement for number of clusters. Moreover the results establish that the starting energy requirement is dependent on number of clusters and the network diameter from the base station. It has also been pragmatic that the starting/initial energy requirement is dependent upon the data frames used. It must be kept highest when number of clusters is less and network diameter is more.

**Manju Sharma** received her B.Tech from UP Technical University, Lacknow, India and Master of Technology in Computer Science and Engineering from Punjab Technical University Jalandhar, in the year 2007. From 2004 she is working as a Lecturer in Information Technology at DAV Institute of Engineering and Technology, Jalandhar. Currently she is working in the area of data communication, computer network and wireless sensor networks and pursuing her Ph.D. She has published 12 research papers in the International/National/Conferences. She is member of Punjab Science Congress, Patiala, India

**Lalit K Awasthi** born on May 19, 1966 at Sunder Nagar (Himachal Pardesh, India). He received his BE degree in Computer Science & Engineering from Shivaji University in 1988. He joined as Lecturer in Department of Computer Science and Engineering at Regional Engineering College, Hamirpur, India on August 1988. He then received his M. Tech., Computer Science and Engineering degree from Indian Institute of Technology, Delhi in 1993. He joined CSE department, REC Hamirpur, India as Assistant Professor on 15th September 1994. He received his Ph.D. degree in Computer Technology from Indian Institute of Technology Roorkee in 2003 and joined as Professor at the same NIT, Hamirpur, India in July 2004.

He has published 50 papers in International and National Journals and Conferences. He has guided many B. Tech. and M. Tech. thesis. His areas of interest are Computer Architecture, Fault Tolerant, Parallel & Distributed Processing, Checkpointing, Mobile Computing, Ad hoc and Sensor networks. He is guiding ten research scholars. Presently, he is working as Head, Computer Centre at National Institute of Technology, Hamirpur, India since August 2009. Earlier he had served as Head of Computer Science and Engineering Department from Jan 1998 to September 1999 and from October 2002 to February 2006.